\documentclass[pdflatex,bst/sn-mathphys-num]{sn-jnl}

\usepackage{graphicx}%
\usepackage{multirow}%
\usepackage{amsmath,amssymb,amsfonts}%
\usepackage{amsthm}%
\usepackage{mathrsfs}%
\usepackage[title]{appendix}%
\usepackage{xcolor}%
\usepackage{textcomp}%
\usepackage{manyfoot}%
\usepackage{booktabs}%
\usepackage{algorithm}%
\usepackage{algorithmicx}%
\usepackage{algpseudocode}%
\usepackage{listings}%
\usepackage{subcaption} % No preâmbulo
\DeclareMathSymbol{\Tau}{\mathalpha}{operators}{"54}

\theoremstyle{thmstyleone}%

\theoremstyle{thmstyletwo}%

\theoremstyle{thmstylethree}%

\begin{document}

\title[Article Title]{Evolution of density variations in a trapped atomic superfluid driven into turbulence}

% Texture evolution during the development of turbulence in a trapped atomic superfluid
% Density Variations During the Development of Turbulence in a Trapped Atomic Superfluid
% Evolution of Density Variations in a Trapped Atomic Superfluid Driven into Turbulence
% Decay of Density Variations Following Turbulent Excitation in a Trapped Atomic Superfluid

\author*[1]{\fnm{Michelle A.} \sur{Moreno-Armijos}}\email{michelle.moreno@ifsc.usp.br}

\author[1]{\fnm{Leandro} \sur{Machado}}\email{lemachado@usp.br}

\author[1]{\fnm{Arnol D.} \sur{García-Orozco}}\email{arnolgarcia@ifsc.usp.br}

\author[1]{\fnm{Sarah} \sur{Sab}}\email{sarahsab@usp.br}

\author[1]{\fnm{Amilson R.} \sur{Fritsch}}\email{amilson@usp.br}

\author*[1,2]{\fnm{Vanderlei S.} \sur{Bagnato}}\email{vander@ifsc.usp.br}

\affil[1]{\orgdiv{Sao Carlos Institute of Physics}, \orgname{University of Sao Paulo, IFSC - USP}, \postcode{13566-590}, \orgaddress{\city{Sao Carlos}, \state{SP}, \country{Brazil}}}

\affil[2]{\orgdiv{Department of Biomedical Engineering}, \orgname{Texas A\&M University}, \orgaddress{\city{College Station}, \postcode{77843}, \state{Texas}, \country{USA}}}

\abstract{

In this work, we study the free decay of a turbulent trapped Bose gas by analyzing the temporal evolution of density variations extracted from absorption images. We introduce a parameter $\delta$ as a simple and experimentally accessible observable that captures the amplitude of density variations. After the driving is turned off, this parameter exhibits a clear decay, which enabled us to identify a characteristic relaxation time. Interestingly, this timescale remains nearly constant across the range of excitation amplitudes explored, while the magnitude of $\delta$ varies with the injected energy. Numerical simulations based on the Gross-Pitaevskii equation reveal a qualitatively similar behavior, both showing a decay of density variations over time.
}

\keywords{Quantum turbulence, Density variations, Relaxation of turbulence}



\maketitle

\section{Introduction}\label{sec1}

Studying how many-body systems evolve far from equilibrium allows for a deeper understanding of fundamental principles, such as thermalization and equilibration, related to emergent phenomena, including topological phases of matter \cite{mcginley2018topology}, quantum chaos, many-body localization \cite{rispoli2019quantum} and non-equilibrium phase transitions \cite{eisert2015quantum}. To advance the understanding of these complex many-body systems it becomes essential to develop new tools and techniques that complement the existing approaches for characterizing those systems \cite{polkovnikov2011colloquium}. In this work, we aim to study the evolution of the amplitude of density variations during the establishment and relaxation of a turbulent Bose-Einstein condensate (BEC). Examples of studies on density variation distributions include the analysis of density patterns in different materials and systems, where such variations can affect mechanical, thermal, or optical properties. Research on these density patterns has been carried out in various physical contexts, including magnetic materials \cite{muhlbauer2009skyrmion}, liquid crystals \cite{dierking2003textures}, or models of the early Universe \cite{vilenkin1981gravitational}. The development of experimental techniques involving BECs has enabled the study of spin textures, which serve as analogs to the systems mentioned above \cite{hong2023spin,ruben2010texture}. 
Correlations of density fluctuations have also been extensively studied theoretically \cite{naraschewski1999spatial,henseler2008density,kohnen2015temporal} and experimentally \cite{jeltes2007comparison,perrin2012hanbury,hertkorn2021density} in several atomic systems, including BECs. However, the experimental application of this approach remains challenging due to the requirement for high-resolution detection systems, and the need for a large number of measurements owing to the statistical nature of fluctuations \cite{folling2014quantum}. For this reason, we present a simplified method for studying the evolution of the magnitude of density variations in the momentum distribution of a turbulent BEC, which, although with limitations, can reveal intrinsic properties of out-of-equilibrium quantum systems. A similar approach, presented by Nagler et al. \cite{nagler2022observing}, has been used to study the response of density fluctuations and phase coherence in BECs through disorder quenches.

Turbulence in quantum gases usually starts with a condensate in which excitations such as vortices, solitons, and waves are introduced by applying external perturbative fields \cite{middleton2023strong, barenghi2023types}. Once these perturbations are established, the mutual interactions and reactions lead to the development of turbulence. As turbulence progresses, energy typically moves from large to small scales, causing variations in density as the energy distribution evolves among scales. A key challenge is to identify observables that can reliably trace the evolution of turbulence, so in this work, we employ a simple approach to estimate a decay time of density variations related to the relaxation dynamics of a turbulent BEC.

%A key challenge is identifying measurable quantities that track the evolution and establishment of turbulence over time. Here, we used a simple method to quantify this evolution, enabling the monitoring of the dynamics of a turbulent atomic BEC during its relaxation. %These density variations are significantly influenced by interactions and external potentials. 

In our experiment, the primary observable is the momentum distribution  $n(k, t_\mathrm{hold})$, where $k$ denotes the momentum and $t_\mathrm{hold}$ is the time elapsed since relaxation began, following the end of the excitation. Using absorption imaging in time of flight (TOF), we analyze variations in the momentum distribution, revealing distinct evolution patterns that depend on the amount of energy injected into the system. In addition, we present numerical simulations of the turbulent system modeled by the Gross-Pitaevskii equation (GPE). The numerical simulations of \textit{in situ} clouds were sufficient to observe the evolution of density variations patterns during relaxation, and the findings are consistent with the experimental measurements obtained after TOF expansion.

\section{Tracking density variations via absorption imaging}

% Absorption imaging is a widely used technique in cold atom experiments to measure the density distribution of atomic clouds. Sin embargo, para obtener la distribución de momento del sistema usando imagen de absorción después de expansión en TOF, el sistema necesariamente debe ser dominado cinéticamente. Aunque, por nuestras limitaciones experimentales esto es difícil de comprobar en nuestro sistema, en estudios previos de nuestro grupo \cite{caracanhas2013self,thompson2013evidence} hemos demostrado computacionalmente que the shape of the expanded turbulent cloud is mainly determined by the polarization of the \textit{in situ} vortex tangle, resulting in a kinetic energy per particle that significantly exceeds the mean-field interaction energy. Esta técnica ha sido usada ampliamente en varios trabajos de nuestro grupo \cite{thompson2013evidence, garcia2020intrascales, garcia2022universal, moreno2025observation}, así como en varios trabajos de otros grupos \cite{navon2016emergence}. In simulaciones de nuestro experimento \cite{middleton2023strong}, se ha evidenciado que turbulence is induced by injecting energy through displacements and rotations, creating a nonequilibrium condensate with strong density and phase fluctuations caused by vortex reconnections, por lo tanto, podemos considerar que after expansion nuestro sistema in fact is kinetically dominated and the absorption images reflect the momentum distribution. \\

Absorption imaging is a widely used technique in cold atom experiments to measure the density distribution of atomic clouds. However, to reliably extract the momentum distribution after TOF expansion, the system must be kinetically dominated. Although our experimental limitations make this difficult to verify in our system, previous studies from our group \cite{caracanhas2012self,caracanhas2013self, thompson2013evidence} have shown numerically that the shape of the expanded turbulent cloud is determined by the polarization of the \textit{in situ} vortex tangle, resulting in a kinetic energy per particle that significantly exceeds the mean-field interaction energy. The absorption imaging technique has been extensively employed in several of our works \cite{thompson2013evidence, garcia2020intrascales, garcia2022universal, moreno2025observation}, and in various studies reported in the literature \cite{navon2016emergence,fabbri2011momentum}. Simulations of our experimental protocol \cite{middleton2023strong} confirm that turbulence is generated through energy injection through displacements and rotations, leading to a nonequilibrium condensate with strong density and phase fluctuations due to vortex reconnections. Therefore, we may assume that after expansion, the system is kinetically dominated, and the absorption images reflect the momentum distribution.

% Since TOF measurements represent the momentum distribution $n(k)$, we refer to it in momentum space.
The absorption imaging projects the three-dimensional momentum distribution onto a plane perpendicular to the propagation direction of the probe beam. This projection assigns a value to each pair $(k_y, k_z)$, denoted as $n_\mathrm{m}(k_y, k_z)$. Considering an atomic BEC, it is well known that it exhibits a characteristic smooth profile in equilibrium, which becomes significantly disturbed when excitations introduce disorder. To quantify the amplitude of density variations while relaxation takes place, we propose a simple approach derived from the residuals of fitting the two-dimensional momentum distribution to a function that better describes the cloud in equilibrium. The bimodal function consisting of Thomas-Fermi and Gaussian components is widely accepted to describe a BEC in equilibrium in a harmonic trap \cite{pethick2008bose,ketterle1999making}, and it is given by 
\begin{equation}
\label{eq:bimodal}
\begin{split}
        n_\mathrm{fit}(k_y,k_z) = A_\mathrm{TF} \, \max & \left[ \left( 1 -  \frac{(k_y - k_{y0})^2}{R_y^2} - \frac{(k_z - k_{z0})^2}{R_z^2} \right)^{3/2}, 0 \right] \\ 
        & + A_\mathrm{G} \exp \left[ -\frac{(k_y - k_{y0})^2}{\sigma_y^2} - \frac{(k_z - k_{z0})^2}{\sigma_z^2} \right],
\end{split}
\end{equation}
where $A_\mathrm{TF}$ and $A_\mathrm{G}$ are the amplitudes of the column-integrated Thomas-Fermi and two-dimensional Gaussian functions, respectively, $k_{y0}$ and $k_{z0}$ are the coordinates of the center of mass, $R_y$ and $R_z$ are the Thomas-Fermi radii, and $\sigma_y$ and $\sigma_z$ are the Gaussian widths in momentum space. Inspired by the definition of root mean square deviation, we introduce the density variations parameter $\delta$ as
\begin{equation}
\label{eq:delta}
\delta = \sqrt{ \left\langle \left[ n_\mathrm{m}(k_y,k_z) - n_\mathrm{fit}(k_y,k_z) \right]^2 \right\rangle },
\end{equation}
where the brackets denote averaging over all pixels of the absorption image. This means that the projection of the density in the $k_y k_z$ plane, can be divided into several elements positioned in $(k_{yi}, k_{zj})$ whose values of $n_\mathrm{m}(k_{yi}, k_{zj})$ and $n_\mathrm{fit}(k_{yi}, k_{zj})$ are determined. Unlike the calculus of residuals, where the residual values are obtained by the differences at each point, we want to obtain an average value representative of those fluctuations in the density distribution. By computing the mean value in Equation \ref{eq:delta}, we avoid the contribution of extreme or outlier values of density that can significantly bias the interpretation of the results. 
The evolution of $\delta$ exhibits the distribution and amplitude of the fluctuations relative to the fitting. Since the number of particles is conserved, the integral $ \int n(k_y,k_z) \, \mathrm{d}k_y \, \mathrm{d}k_z $ is constant regardless of the size of perturbations, or instant of time. In this sense, $\delta$ describes the density variations of the momentum distribution profile, as larger values correspond to greater deviations and fluctuations. In contrast, smaller values correspond to a larger number of variations but of lower amplitude. 

\section{Experiment}

We begin with a rubidium-87 BEC, in the hyperfine state $F=1$ and $m_F = -1$,  magnetically trapped in a harmonic potential characterized by radial and axial frequencies of $\omega_r/2\pi = 110.9(2)$ Hz and $\omega_x/2\pi = 15(1)$ Hz, respectively. The BEC contains approximately $N \approx 2 \times 10^5$ atoms and a condensed fraction $N_0/N$ above 80\%. We introduce excitations using an external oscillatory magnetic field to drive the system to a turbulent state. This perturbation is generated by a pair of coils arranged in anti-Helmholtz configuration, with one of the coils being displaced along the $z-$ and $y-$axis, and tilted at a small angle of about $5^\circ$ respect to the $x-$axis. An oscillatory electric current passing through the coils creates a shifting and compressing potential at the center of the trap. The excitation amplitude $A$ (amplitude of the excitation potential) is expressed in terms of the chemical potential $\mu_0$. We vary the excitation amplitude while keeping the excitation time fixed at $t_{\text{exc}} = 8\mathrm{T}$ (where $\mathrm{T}$ is the excitation period) and the excitation frequency at $\omega_{\text{exc}} = 2\pi/\mathrm{T} = 2\pi \times 105$ Hz. This corresponds to $\omega_{\text{exc}} \approx 0.95\, \omega_r$, ensuring efficient coupling to the dipole mode. Although the trap is symmetric in the $y$ and $z$ directions, this excitation predominantly couples the dipole mode along the $z$ direction. In addition, we also observe shape deformations of the cloud that we associate with a coupling of quadrupolar and breathing modes that oscillate with a frequency of approximately $2\omega_r$. However, all measurements are taken as the condensate crosses the trap center (where kinetic energy is maximal), ensuring that the aspect of the cloud remains consistent across cycles, allowing us to disregard the influence of shape deformations (see Appendix A).

Once the excitation stops, we allow the atomic cloud to relax in the trap during a holding time $t_\mathrm{hold}$, then we probe its state using absorption imaging after 30 ms of TOF expansion. The imaging beam is incident on the sample perpendicularly to the $z-$axis and at a $45^{\circ}-$angle with respect to the $y-$axis. From the captured images, we obtain the momentum distribution $n_\mathrm{m}(k_{xy},k_z)$, where $k_{xy}$ comes from the projection of the momentum with contributions in $x-$ and $y-$axis due to the direction of the imaging beam, i.e., $\textbf{k}_{xy} = k_{x}\sin{(45^\circ)} \hat{e}_x + k_{y}\cos{(45^\circ)} \hat{e}_y$. The two-dimensional momentum distributions are normalized such that $2\pi \int n_\mathrm{m}(k_{xy}, k_z) \, \mathrm{d}k_{xy} \, \mathrm{d}k_z = 1$. The total number of atoms is conserved throughout the dynamics, remaining approximately constant over all considered times.

\begin{figure}[ht]
\centering
\includegraphics[width=\textwidth]{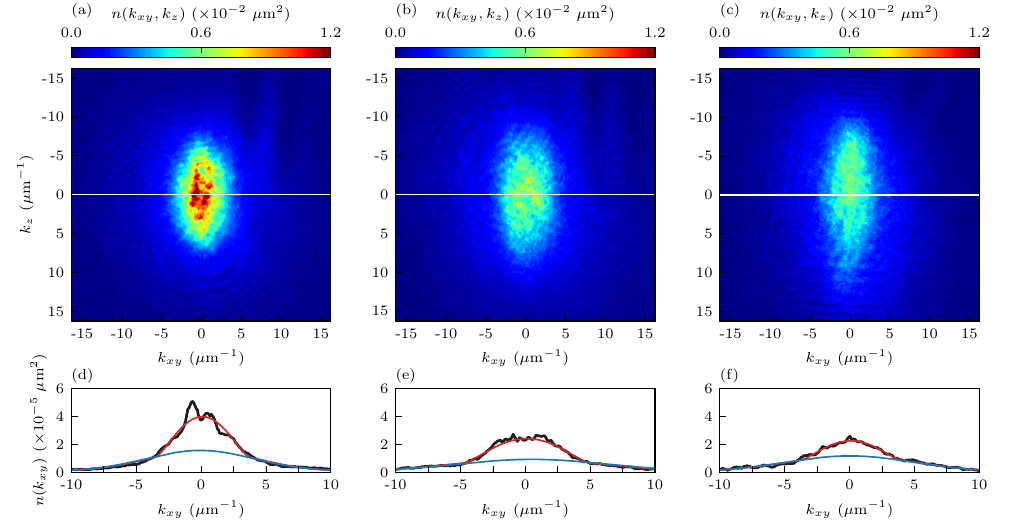}
\caption{Momentum distribution of an excited BEC. (a-c) Two-dimensional integrated momentum distribution obtained via absorption imaging for three excitation amplitudes: $1.2\mu_0$, $1.6\mu_0$, and $2.0\mu_0$, respectively. In all cases, the hold time is $t_\mathrm{hold}\approx 100$ ms. (d-f) One-dimensional slices of the momentum distribution (black curves) are taken through the center, along the $xy-$axis, as indicated by the horizontal white lines in figures (a-c). The red curves correspond to a slice of the two-dimensional bimodal fitting of Equation \ref{eq:bimodal}, while the blue curves indicate the Gaussian component of the fit. These profiles show the change in the level of density variations depending on the injected energy.}
\label{fig:fig1}
\end{figure}

In a previous work \cite{garcia2022universal}, we characterized the onset of the turbulent state as a function of the excitation amplitude by identifying a power-law behavior in the momentum distribution, indicating an energy cascade. The cascade emerges after the excitation is turned off, as the system requires a finite relaxation time to redistribute the injected energy among scales to establish the cascade. In this case, we found that this process takes approximately 20 ms, and for an excitation amplitude of $A = 1.2\mu_0$, the system begins to exhibit an energy cascade. Figure \ref{fig:fig1} shows examples of the momentum distribution of the excited BEC with excitation amplitudes $1.2\mu_0$, $1.6\mu_0$, and $2.0\mu_0$ after a holding time $t_\mathrm{hold}\approx~100$ ms. The top panels display absorption images, showing the normalized two-dimensional momentum distribution and evidencing how density variations change with increasing excitation amplitude. In the bottom panels, the black curves represent a one-dimensional slice of the momentum distribution taken through the center of the cloud along the $y-$axis. The solid red curves correspond to a slice of the two-dimensional bimodal fitting, described in Equation \ref{eq:bimodal}, while the blue curves correspond to the Gaussian component of the fit. These examples highlight the emergence of large-scale perturbations that gradually transform into smaller-scale features, altering the amplitude of the density variations in the momentum distribution.\\

\begin{figure}[ht]
\centering
\includegraphics[width=1.04\textwidth]{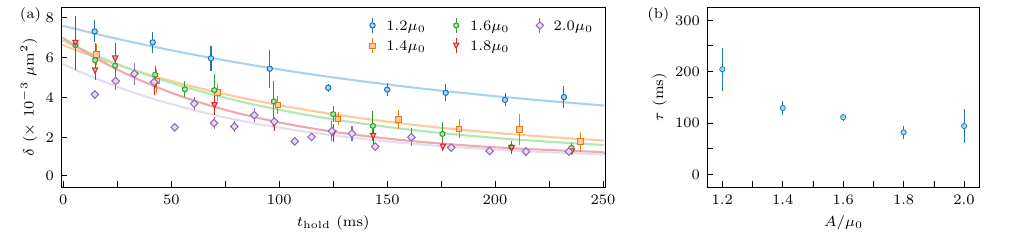}
\caption{(a) Evolution of the density variations parameter $\delta$ as a function of hold time, measured experimentally for different excitation amplitudes. Error bars indicate the standard deviation derived from repeated measurements. We fitted an exponential function (solid curves) based on the observed decay. (b) The decay time $\tau$, extracted from the fitting, is plotted as a function of the excitation amplitude. The error bars represent the uncertainties associated with the fitting results.}
\label{fig:fig2}
\end{figure}

We characterize the evolution of density variations by measuring the deviation from the bimodal profile applying Equation \ref{eq:delta}. A dataset of the measured $\delta$ for different excitation amplitudes is shown in Figure~\hyperref[fig:fig2]{\ref{fig:fig2}(a)}; each data point represents a value of $ \delta $ at a specific time during the relaxation process. The curves indicate the results of fitting the data to an exponential decay function of the form $ \delta(t_\mathrm{hold}) = \delta_0\, e^{-t_\mathrm{hold}/\tau} + \text{offset} $, where $ \tau $ represents the characteristic decay time, and $ \delta_0 $ the amplitude of the decay. It is important to mention that disorder starts to develop during the excitation time $t_\mathrm{exc} = 76$ ms, leading to different values of $\delta_0$ depending on the excitation amplitude. Within the range of excitation amplitudes considered in Figure~\hyperref[fig:fig2]{\ref{fig:fig2}(a)}, we observe that high amplitudes show initial values of $\delta$ smaller than for lower amplitudes, indicating that density variations are initially less pronounced due to a faster evolution to a finer distribution of density variations while still exciting. In contrast, small excitation amplitudes lead to the coarsening of density variations, resulting in larger values of $\delta$. Eventually, the curves converge to similar values of $\delta$, characterizing the distribution of density variations in the final stage of evolution on the path to thermalization. It is important to remark that in equilibrium, $\delta$ remains nonzero due to intrinsic experimental fluctuations. However, since the system is not kinetically dominated in this regime, absorption images do not accurately reflect the momentum distribution. For this reason, we do not include $\delta$ curves for the equilibrium case.
\\

The decay of $\delta$ reflects the speed at which the excitations interact, producing a progressively finer distribution of density fluctuations. Figure \hyperref[fig:fig2]{\ref{fig:fig2}(b)} shows the values of the decay time $\tau$ as a function of the excitation amplitude $A$, extracted from the fits in Figure \hyperref[fig:fig2]{\ref{fig:fig2}(a)}. Excluding the case of $A =1.2\mu_0$, which corresponds to the onset of the turbulent state and significantly deviates from the decay times observed with other amplitudes, the decay time remains approximately constant with an average value of approximately $104$ ms within this range of excitation amplitudes.

\section{Numerical simulation}

To complement the experimental analysis of density variations, we performed numerical simulations using the three‑dimensional time‑dependent GPE:

\begin{equation}\label{eq:TDGPE}
    i\hbar \frac{\partial\psi(\mathbf{r},t)}{\partial t} = 
\left(-\frac{\hbar^2}{2m}\nabla^2 + V(\mathbf{r},t) + g|\psi(\mathbf{r},t)|^2 - \mu \right)\psi(\mathbf{r},t),
\end{equation}

where $\psi(\mathbf{r},t)$ is the condensate wavefunction with $\mathbf{r} = (x,y,z)$ the spatial coordinate. Here, $\hbar$ is the reduced Planck constant, $m$ is the atomic mass, and $V(\mathbf{r},t)$ is the external trapping and excitation potential.
The interatomic interaction constant is $g = 4\pi\hbar^2 a_s/m$, with $a_s$ the $s-$wave scattering length, and $\mu$ denotes the chemical potential.

The simulation consists in solving Equation \ref{eq:TDGPE} in a cubic grid of $N_x = 352$, $N_y = 128$ and $N_z = 128$ points with a spacing $dl = \xi/2 = 0.125$, where $\xi$ is the healing length. The time integration is performed using the \textit{Four Order Runge-Kutta method} with time spacing of $dt = 2\times 10^{-3}$ while the spatial derivatives were computed using the finite-differences method. Before evolving the system in the real time, we used the \textit{imaginary-time propagation technique} \cite{itime} to prepare the system in its ground state. This technique consists in replacing $t$ by $-it$ so that Equation~\ref{eq:TDGPE} becomes a diffusion-like equation, which exponentially leads the system to the ground-state. 

We considered a BEC of $1.0\times 10^5$ atoms of $^{87}$Rb trapped in the axially symmetric harmonic potential:
\begin{equation}
  V_{\mathrm{trap}}(x,y,z) = \frac{1}{2}m\omega_r^2\left( \lambda^2x^2 + y^2 + z^2 \right),  
\end{equation}
where $\lambda = \omega_x/\omega_r$ is the anisotropy factor. In the experiment, this parameter is around $1/7.3$, giving rise to a cigar-shaped BEC. Besides the confining potential, the experiment holds an extra potential responsible for promoting excitations in the BEC. In this setup, there is a slight tilt between the excitation coils ($\theta_x \approx 5^{\circ}$) and the $y$ axis of the trapping potential. This misalignment is responsible for breaking the symmetry of the system, leading to an excitation potential given by \cite{middleton2023strong}
\begin{equation}
    \label{eq:Holly_Pot}
    V_{\mathrm{exc}}(x,y,t) = A[1-\cos{(\omega_{\mathrm{exc}}t)}]x'/R_x,
\end{equation}
where $R_x=\sqrt{2\mu/m\lambda^2\omega_x^2} $ is the Thomas-Fermi radius along the $x$-direction, $A$ is the excitation amplitude in units of chemical potential that characterizes the excitation strength, and $x' =  \cos{(\theta_x)}x - \sin{(\theta_x)}y$. The total potential considered in the simulation is the contribution of both trap potential and excitation potential, that is,
\begin{equation}
    V(\mathbf{r},t) = V_{\mathrm{trap}}(x,y,z) + V_{\mathrm{exc}}(x,y,t).
\end{equation}
To better represent the experimental conditions, the driving frequency was set at $\omega_{\mathrm{exc}} = 0.95\omega_r$ to enhance the coupling between the external excitation and the system. The simulations were conducted using excitation amplitudes $A/\mu =\{1.0,\, 1.2,\, 1.4,\, 1.6,\, 2.0\}$, applied for a duration corresponding to eight periods, $t_{\mathrm{exc}} = 8\left(2\pi/\omega_{\mathrm{exc}}\right)$. After excitation, we let the system evolve freely in the harmonic trap until the end of the simulation at $t_\mathrm{hold} = 244.8$ ms.\\

Given that the excitation potential of Equation~\ref{eq:Holly_Pot} predominantly acts along the $x-$axis, it generates a periodic force $\mathbf{F}\, =\, -\nabla V_{\mathrm{exc}} = - A[1-\cos{(\omega_\mathrm{exc}t)}]\, \left(\cos{(\theta_x)} \hat{e}_x - \sin{(\theta_x)\hat{e}_y} \right)$, and the dipolar mode along $x$ becomes the most significant collective mode in the system. The amplitude of this mode is substantial, causing the center of mass of the cloud to oscillate around the minimum of the trap potential, thus demanding the use of a large computational grid. To avoid it, we set $\lambda = 1/2$, which allows the use of a smaller grid in the simulation.\\

For low excitation amplitudes ($A < 1.0\mu$), the effect of the small angle between the excitation coils can be neglected, since the excitation is primarily made along the $x$ direction. In contrast, for high amplitudes ($A\geq 1.0\mu$), the misalignment is essential to break the symmetry of the system around the $x$ axis, which is crucial for generation of turbulence, as it induces both translation and rotation by introducing angular momentum and deforming the trap due to the shifted potential minimum. Consequently, the formation of characteristic superfluid topological defects such as vortices and solitons is expected \cite{middleton2023strong}. Higher excitation amplitudes nucleate many vortices modulated by density fluctuations, thereby inducing random dynamics and fostering quantum turbulence.  \\

\begin{figure}[b]
\centering
\includegraphics[scale=1.0]{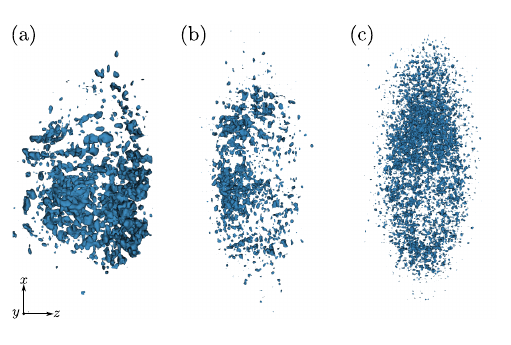}
\caption{Density isosurface plots of fluctuations in the cloud when $A = 1.4\mu$. (a) At $t_\mathrm{hold} = 4.34$ ms just after the excitation protocol, (b) at an intermediate time $t_\mathrm{hold} = 69.45$ ms, and (c) at a late time $t_\mathrm{hold} = 214.14$ ms, illustrating the evolution from a coarse to fine distribution of density variations.}
\label{fig:texclouds}
\end{figure}

Although experiment and simulation show similar overall features, they differ in two key aspects. First, the simulation assumes zero temperature, neglecting the thermal component that is typically present in experiments and coexists with the condensate. Second, in the experiments, images of the cloud are obtained using absorption imaging in TOF. To minimize these discrepancies, the clouds were experimentally prepared with a high condensed fraction (typically $>80\%$), ensuring that the influence of the unavoidable thermal component remains as minimal as possible. Moreover, we perform a Fourier transform of the computed wave function in real space to transition to momentum space. Following this step, the parameter $\delta$ is evaluated as a function of time.\\

Simulations show that the system begins to form fragments with sizes comparable to the scale of the entire BEC \cite{middleton2023strong}, significantly deviating from the equilibrium Thomas-Fermi profile, showing a coarse distribution of density variations, as shown in Figure \hyperref[fig:texclouds]{\ref{fig:texclouds}(a)}. As evolution progresses, the excitations gradually decrease in size [Figure \hyperref[fig:texclouds]{\ref{fig:texclouds}(b)}], and eventually, the system reaches a state characterized by a thin distribution of density variations, where energy populates higher momenta, as shown in Figure \hyperref[fig:texclouds]{\ref{fig:texclouds}(c)}. Based on these observations, a gradual decay of $\delta$ is expected during the hold time, but in the ideal equilibrium scenario, $\delta$ remains zero, indicating agreement with the Thomas-Fermi profile and the absence of density variations. \\

\begin{figure}[b]
\centering
\includegraphics[scale=0.8]{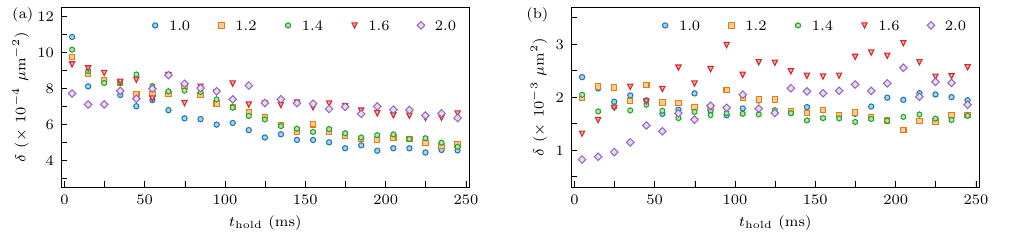}
\caption{ Temporal evolution of $\delta$ for different excitation amplitudes $A$ computed using Equation \ref{eq:delta} in (a) real space and (b) momentum space. In (a), $\delta$ is computed in spatial coordinates. The curves show how the initial excitation amplitude influences the magnitude of density variations and their relaxation towards a quasi-stationary state.}
\label{fig:rspace_kspace}
\end{figure}

The time evolution of $\delta$ in real space is presented in Figure \hyperref[fig:rspace_kspace]{\ref{fig:rspace_kspace}(a)}, and was computed using Equation \ref{eq:delta} in spatial coordinates. For the largest excitation amplitude, $A = 2.0\mu$, $\delta$ initially increases before decaying, this can be explained because the system is still responding to the injected energy and defects continue to form even after the drive ceases. In contrast, lower amplitudes display a  decrease of $\delta$, indicating that these cases rapidly enter a relaxation phase in which interactions between excitations generate progressively finer density patterns. \\

%In momentum space, as shown in Figure \hyperref[fig:rspace_kspace]{\ref{fig:rspace_kspace}(b)}, a similar behavior to the real-space case is observed. However, for $A=1.6\mu$ and $A=2.0\mu$, there is an initial increase in $\delta$ before it decays. This can be attributed to the excitation of higher-momentum modes, which create coarser {\color{blue} distribution of density variations} before the system redistributes energy into finer patterns of density variations. The interpretation for real and momentum space is similar but exhibits subtle nuances. In real space, the {\color{blue}density variations} reflect the spatial redistribution of particles, which, over longer times, tends to evolve into a more regular and uniform pattern characterized by reduced density variations. In momentum space, a finer {\color{blue} distributions of density variations} corresponds to a greater concentration of energy in high-lying modes at later times. In both scenarios, higher amplitudes result in elevated initial values of $\delta$, yet all cases demonstrate a tendency to saturate at a plateau.

In momentum space [Figure \hyperref[fig:rspace_kspace]{\ref{fig:rspace_kspace}(b)}], the initial increasing behavior of $\delta$ for high excitation amplitudes ($A > 1.6\mu$) is more evident. Again, we attribute this effect to the system continuing to respond to the injected energy, with defects forming even after the drive ends. In the long-time regime, $\delta$ in real space [Figure \hyperref[fig:rspace_kspace]{\ref{fig:rspace_kspace}(a)}] decreases for all excitation amplitudes and eventually reaches a plateau, indicating that the system relaxes toward a quasi-stationary state. In momentum space [Figure \hyperref[fig:rspace_kspace]{\ref{fig:rspace_kspace}(b)}], the behavior is similar, although $\delta$ exhibits larger fluctuations and a slight increase for higher excitation amplitudes before settling. In both representations, the asymptotic value of $\delta$ is set by the initial excitation amplitude, making the long-time behavior a qualitative indicator of the residual disorder and degree of non-equilibrium sustained by the system.\\

\begin{figure}[b]
\centering
\includegraphics[scale=0.8]{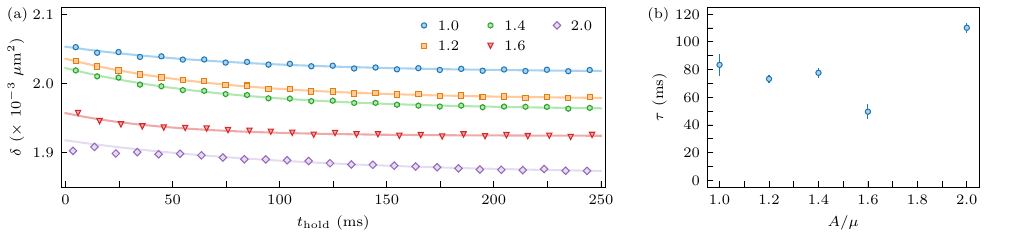}
\caption{(a) Evolution of the density variations parameter} $\delta$ calculated in momentum-space considering a projection of $45^\circ$ respect to the $y-$axis. The solid curves correspond to the same exponential decay fitting used in the previous section. (b) Decay time as a function of the excitation amplitude obtained from fitting the curves of panel (a). The error bars correspond to the uncertainties associated with the fitting results.
\label{fig:kspace45}
\end{figure}

When considering the angular projection of the experimental imaging method, the simulation reveals a distinct behavior. As illustrated in Figure \hyperref[fig:kspace45]{\ref{fig:kspace45}(a)}, higher excitation amplitudes now result in lower values of $\delta$, displaying a trend similar to that observed in the experiment [Figure \hyperref[fig:fig2]{\ref{fig:fig2}(a)}]. This inversion arises because the integration at $45^\circ$ smooths contributions from different directions in the momentum distribution profile, effectively mixing contributions from both more and less confined directions and producing an apparent decrease of $\delta$ for high excitation amplitudes. As the excitation acts along the elongated axis of the BEC, the energy spreads preferentially in that direction; therefore, the angular integration mixes momentum components smoothing anisotropy effects.\\

Figure~\hyperref[fig:kspace45]{\ref{fig:kspace45}(b)} shows the decay time $\tau$ obtained from the curves in Figure~\hyperref[fig:kspace45]{\ref{fig:kspace45}(a)} via an exponential fit. We find that, for high excitation amplitudes ($A > 1.6\mu$) as the system is still reacting to the injected energy at early stages of the hold time, the density variations do not display a clear exponential decay at this stage, making the fitted values of $\tau$ less reliable. However, excluding the cases with $A > 1.6\mu$, the decay time remains approximately constant, yielding an average value of $\tau \approx 79$ ms. Despite the differences between experiment and simulation, the decay times shown in figures \hyperref[fig:fig2]{\ref{fig:fig2}(b)} and \hyperref[fig:kspace45]{\ref{fig:kspace45}(b)} exhibit similar behavior, remaining approximately constant over time. Specifically, the decay time is $\approx 104$ ms in the experiment and $\approx 79$ ms in the simulation. The fact that the decay time in the simulation is shorter than in the experiment, although counterintuitive since thermal dissipation would typically accelerate the decay, may be partially attributed to the lower atom number used in the simulations. However, a more detailed investigation of this discrepancy merits further investigation.

\section{Conclusion}
In this work, we investigated the evolution of density variations in a freely decaying turbulent BEC by analyzing absorption images acquired during relaxation after a controlled excitation was applied. To quantify density variations, we introduced a parameter $\delta$, and studied its temporal behavior for different excitation amplitudes.\\

In a previous study \cite{moreno2025observation}, we showed that the turbulent dynamics of our system can be described within the framework of weak wave turbulence theory in the four-wave mixing regime, where the excitations follow the high-$k$, free-particle limit of the Bogoliubov spectrum. However, the information that can be experimentally obtained when studying such systems is inherently limited, making it essential to maximize the information extracted with the available resources.  Motivated by this, we proposed the analysis of density variations as a simple and experimentally accessible observable that can complement the characterization of the turbulent regime without the need for full spectral reconstruction or detailed momentum-space analysis.\\

Our experimental results show that $\delta$ exhibits a clear decay behavior after the excitation is turned off, allowing us to identify a characteristic decay time of approximately $\tau \approx 104$ ms. Remarkably, for the range of excitation amplitudes considered, this decay time remains nearly constant, whereas only the magnitude of density variations depends on the excitation amplitude. This indicates that the relaxation dynamics of the system, once turbulence is established, follow a timescale that is largely independent of the excitation strength. Overall, the decay of $\delta$ reflects the transition from an initially coarse distribution of density variations to increasingly finer and more homogeneous structures as energy is redistributed internally. We interpret this decay time as a signature of the redistribution of excitations in the condensate and the equilibration process. This interpretation is further supported by recent experimental observations from our group, where a similar equilibration time was found during the establishment of different types of turbulent cascades \cite{moreno2025observation}.\\

However, a notable discrepancy remains between the experimental results and simulations, which may stem from several factors. One of the most significant is the presence of a thermal cloud, whose role in the dynamics is not yet fully understood. In the experiment, the thermal atom fraction (20\%) can readily occupy low-density regions of excitations, causing the system to behave differently than it would at zero temperature. Another factor is the challenge of correlating TOF measurements with the \textit{in situ} momentum distribution. Ensuring that the energy of the sample is predominantly kinetic is difficult to verify without additional evidence of invariance under fixed experimental conditions. Despite these limitations, our observations support the validity of using density variations as a parameter to monitor the temporal evolution of a turbulent trapped superfluid. 
Undoubtedly, a deeper understanding of these findings will require more accurate simulations and continued theoretical and experimental investigation in future work.

\bmhead{Acknowledgements}
The authors thank L. Madeira and M. Caracanhas for fruitful discussions. This work was supported by the São Paulo Research Foundation (FAPESP) under Grants No. 2013/07276-1 and No. 2014/50857-8, by the National Council for Scientific and Technological Development (CNPq) under Grants No. 465360/2014-9. M.A.M-A. acknowledges the support from Coordenação de Aperfeiçoamento de Pessoal de Nível Superior-Brasil (CAPES)-Finance code No. 88887.643259/2021-00. A.R.F., L.M., A.D.G-O and S.S acknowledge the support from FAPESP-Finance codes No. 2024/08433-8, No. 2024/21658-9, No. 2024/20641-5., No. 2025/07547-2 and No. 2024/14764-7. V.S.B. acknowledges Texas A\&M University (GURI-Governor’s University Research Initiative-M230930).

\bmhead{Competing Interests}
The authors declare no competing interests. 

\bmhead{Author contributions}
M.A.M-A. and A.R.F. collected the data. L.M. performed the numerical simulations. M.A.M-A., A.D.G-O., S.S., and A.R.F. analyzed the data. V.S.B. supervised the project. All authors contributed extensively to interpreting the results and writing the manuscript.

\bmhead{Data availability}
Data is available on reasonable request and should be addressed to the corresponding authors. 

\begin{appendices}

\section{Supplemental material}\label{secA1}

\subsection{Experiment}

\begin{figure}[ht]
\centering
\includegraphics[width=0.9\textwidth]{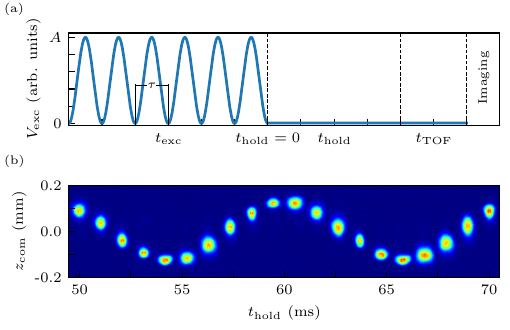}
\caption{(a) Representation of the excitation protocol: after evaporation, an oscillating magnetic is applied with frequency $\omega_\mathrm{exc}$, during $t_\mathrm{exc}$. Then, the system evolves in trap during a holding time, after which the cloud is released, and absorption images are taken after $t_{\mathrm{TOF}}$. (b) Absorption images showing the center-of-mass position $z_{\rm com}$ along the $z-$axis in the time interval $50<t_\mathrm{hold}<70$ ms, evidencing the dipolar and a fast quadrupolar mode oscillating with approximatelly $2\omega_r$.}
\label{fig:supfig1}
\end{figure}

The excitation potential is generated by a pair of coils arranged in an anti-Helmholtz configuration, where one of the coils is displaced along the $z-$ and $y-$axis, and tilted at a small angle of about $5^\circ$ respect to the $x-$axis. An alternating current applied to these coils produces a time-dependent magnetic field that modulates the trapping potential, inducing shifts and compressions of the atomic cloud. This excitation primarily injects energy at a momentum scale on the order of the inverse Thomas-Fermi radius, $k \sim 1/R_{\mathrm{TF}} \approx 0.25\ \mu\mathrm{m}^{-1}$. Figure~\hyperref[fig:supfig1]{\ref{fig:supfig1}(a)} illustrates the timing sequence of the excitation protocol. For all excitation amplitudes, the number of atoms is monitored over time to ensure that remains approximately constant.\\

Key parameters such as amplitude, frequency, phase, and duration are adjustable to explore various excitation regimes. In this work, the excitation period $\tau$ was fixed, with the excitation time given by $t_{\mathrm{exc}} = n \tau$, where $n$ is an integer, and the excitation frequency $\omega_{\mathrm{exc}} = 2\pi / \tau$. We have set $\omega_{\mathrm{exc}} = 2\pi \times 105 $ Hz with $t_{\mathrm{exc}} = 8 \tau$. After the excitation phase, we hold the condensate in the trap for a specific time, referred to as hold time $t_\mathrm{hold}$. To track this evolution, the trap is switched off and absorption imaging is performed after a ballistic expansion with a typical time of flight of $t_{\text{TOF}} = 30$ ms.\\

To calibrate the amplitude of the excitation potential $A$ (which we refer to as excitation amplitude), we apply a fixed current pulse immediately after releasing the cloud and measure the resulting center-of-mass velocity increase $\Delta v = d / t_{\mathrm{TOF}}$, where $d$ is the displacement during time of flight $t_{\mathrm{TOF}}$. From this, the potential variation is estimated as $A \equiv\Delta U = m \, 2 R_{\mathrm{TF}} \, d / (\Delta t \, t_{\mathrm{TOF}})$. Expressed relative to the equilibrium chemical potential at the cloud center $\mu_0$, this quantity defines the excitation amplitude $A$ used throughout this work. 

Despite the trap being symmetric in the $y$ and $z$ directions, the energy coupling predominantly occurs along the $z$ axis, resulting in excitation mainly through the dipole mode. Due to the trap’s characteristics and cylindrical symmetry, other collective modes such as the quadrupolar $(l=2, m=0)$ and breathing $(l=0)$ modes can also be excited. As shown in Figure~\hyperref[fig:supfig1]{\ref{fig:supfig1}(b)}, what we observe in the experiment are shape deformations that oscillate with a frequency approximately $2\omega_r$ that we associate with a coupling of the
quadrupolar and breathing modes, so the dynamics of the system extend beyond pure dipole motion. In addition, we also observe a weak coupling of the scissors mode, but we disregarded it as it does not involve shape deformations. Importantly, all measurements are taken when the condensate crosses the trap center along the dipole trajectory, that is, at integer multiples of half the radial oscillation period, where kinetic energy is maximized and potential energy minimized. As a reference, this point is indicated in Figure~\hyperref[fig:supfig1]{\ref{fig:supfig1}(b)} as the center-of-mass position in the vertical direction $z_\mathrm{com}=0$. This choice ensures that the absorption images after TOF reflect a consistent aspect of the cloud throughout all the measurements, allowing us to disregard density oscillations associated with other modes in our analysis.




\end{appendices}

\bibliography{sn-bibliography}


\begin{thebibliography}{30}
\ifx \bisbn   \undefined \def \bisbn  #1{ISBN #1}\fi
\ifx \binits  \undefined \def \binits#1{#1}\fi
\ifx \bauthor  \undefined \def \bauthor#1{#1}\fi
\ifx \batitle  \undefined \def \batitle#1{#1}\fi
\ifx \bjtitle  \undefined \def \bjtitle#1{#1}\fi
\ifx \bvolume  \undefined \def \bvolume#1{\textbf{#1}}\fi
\ifx \byear  \undefined \def \byear#1{#1}\fi
\ifx \bissue  \undefined \def \bissue#1{#1}\fi
\ifx \bfpage  \undefined \def \bfpage#1{#1}\fi
\ifx \blpage  \undefined \def \blpage #1{#1}\fi
\ifx \burl  \undefined \def \burl#1{\textsf{#1}}\fi
\ifx \doiurl  \undefined \def \doiurl#1{\url{https://doi.org/#1}}\fi
\ifx \betal  \undefined \def \betal{\textit{et al.}}\fi
\ifx \binstitute  \undefined \def \binstitute#1{#1}\fi
\ifx \binstitutionaled  \undefined \def \binstitutionaled#1{#1}\fi
\ifx \bctitle  \undefined \def \bctitle#1{#1}\fi
\ifx \beditor  \undefined \def \beditor#1{#1}\fi
\ifx \bpublisher  \undefined \def \bpublisher#1{#1}\fi
\ifx \bbtitle  \undefined \def \bbtitle#1{#1}\fi
\ifx \bedition  \undefined \def \bedition#1{#1}\fi
\ifx \bseriesno  \undefined \def \bseriesno#1{#1}\fi
\ifx \blocation  \undefined \def \blocation#1{#1}\fi
\ifx \bsertitle  \undefined \def \bsertitle#1{#1}\fi
\ifx \bsnm \undefined \def \bsnm#1{#1}\fi
\ifx \bsuffix \undefined \def \bsuffix#1{#1}\fi
\ifx \bparticle \undefined \def \bparticle#1{#1}\fi
\ifx \barticle \undefined \def \barticle#1{#1}\fi
\bibcommenthead
\ifx \bconfdate \undefined \def \bconfdate #1{#1}\fi
\ifx \botherref \undefined \def \botherref #1{#1}\fi
\ifx \url \undefined \def \url#1{\textsf{#1}}\fi
\ifx \bchapter \undefined \def \bchapter#1{#1}\fi
\ifx \bbook \undefined \def \bbook#1{#1}\fi
\ifx \bcomment \undefined \def \bcomment#1{#1}\fi
\ifx \oauthor \undefined \def \oauthor#1{#1}\fi
\ifx \citeauthoryear \undefined \def \citeauthoryear#1{#1}\fi
\ifx \endbibitem  \undefined \def \endbibitem {}\fi
\ifx \bconflocation  \undefined \def \bconflocation#1{#1}\fi
\ifx \arxivurl  \undefined \def \arxivurl#1{\textsf{#1}}\fi
\csname PreBibitemsHook\endcsname

%%% 1
\bibitem[\protect\citeauthoryear{McGinley and
  Cooper}{2018}]{mcginley2018topology}
\begin{barticle}
\bauthor{\bsnm{McGinley}, \binits{M.}},
\bauthor{\bsnm{Cooper}, \binits{N.R.}}:
\batitle{Topology of one-dimensional quantum systems out of equilibrium}.
\bjtitle{Phys. Rev. Lett.}
\bvolume{121},
\bfpage{090401}
(\byear{2018})
\doiurl{10.1103/PhysRevLett.121.090401}
\end{barticle}
\endbibitem

%%% 2
\bibitem[\protect\citeauthoryear{Rispoli et~al.}{2019}]{rispoli2019quantum}
\begin{barticle}
\bauthor{\bsnm{Rispoli}, \binits{M.}},
\bauthor{\bsnm{Lukin}, \binits{A.}},
\bauthor{\bsnm{Schittko}, \binits{R.}},
\bauthor{\bsnm{Kim}, \binits{S.}},
\bauthor{\bsnm{Tai}, \binits{M.E.}},
\bauthor{\bsnm{L{\'e}onard}, \binits{J.}},
\bauthor{\bsnm{Greiner}, \binits{M.}}:
\batitle{Quantum critical behaviour at the many-body localization transition}.
\bjtitle{Nature}
\bvolume{573}(\bissue{7774}),
\bfpage{385}--\blpage{389}
(\byear{2019})
\doiurl{10.1038/s41586-019-1527-2}
\end{barticle}
\endbibitem

%%% 3
\bibitem[\protect\citeauthoryear{Eisert et~al.}{2015}]{eisert2015quantum}
\begin{barticle}
\bauthor{\bsnm{Eisert}, \binits{J.}},
\bauthor{\bsnm{Friesdorf}, \binits{M.}},
\bauthor{\bsnm{Gogolin}, \binits{C.}}:
\batitle{Quantum many-body systems out of equilibrium}.
\bjtitle{Nature Physics}
\bvolume{11}(\bissue{2}),
\bfpage{124}--\blpage{130}
(\byear{2015})
\doiurl{10.1038/nphys3215}
\end{barticle}
\endbibitem

%%% 4
\bibitem[\protect\citeauthoryear{Polkovnikov
  et~al.}{2011}]{polkovnikov2011colloquium}
\begin{barticle}
\bauthor{\bsnm{Polkovnikov}, \binits{A.}},
\bauthor{\bsnm{Sengupta}, \binits{K.}},
\bauthor{\bsnm{Silva}, \binits{A.}},
\bauthor{\bsnm{Vengalattore}, \binits{M.}}:
\batitle{Colloquium: Nonequilibrium dynamics of closed interacting quantum
  systems}.
\bjtitle{Rev. Mod. Phys.}
\bvolume{83},
\bfpage{863}--\blpage{883}
(\byear{2011})
\doiurl{10.1103/RevModPhys.83.863}
\end{barticle}
\endbibitem

%%% 5
\bibitem[\protect\citeauthoryear{Mühlbauer
  et~al.}{2009}]{muhlbauer2009skyrmion}
\begin{barticle}
\bauthor{\bsnm{Mühlbauer}, \binits{S.}},
\bauthor{\bsnm{Binz}, \binits{B.}},
\bauthor{\bsnm{Jonietz}, \binits{F.}},
\bauthor{\bsnm{Pfleiderer}, \binits{C.}},
\bauthor{\bsnm{Rosch}, \binits{A.}},
\bauthor{\bsnm{Neubauer}, \binits{A.}},
\bauthor{\bsnm{Georgii}, \binits{R.}},
\bauthor{\bsnm{Böni}, \binits{P.}}:
\batitle{Skyrmion lattice in a chiral magnet}.
\bjtitle{Science}
\bvolume{323}(\bissue{5916}),
\bfpage{915}--\blpage{919}
(\byear{2009})
\doiurl{10.1126/science.1166767}
\end{barticle}
\endbibitem

%%% 6
\bibitem[\protect\citeauthoryear{Dierking}{2003}]{dierking2003textures}
\begin{bbook}
\bauthor{\bsnm{Dierking}, \binits{I.}}:
\bbtitle{Textures of Liquid Crystals}.
\bpublisher{Wiley-VCH Verlag},
\blocation{Weinheim}
(\byear{2003}).
\doiurl{10.1002/3527602054} .
\burl{https://onlinelibrary.wiley.com/doi/book/10.1002/3527602054}
\end{bbook}
\endbibitem

%%% 7
\bibitem[\protect\citeauthoryear{Vilenkin}{1981}]{vilenkin1981gravitational}
\begin{barticle}
\bauthor{\bsnm{Vilenkin}, \binits{A.}}:
\batitle{Gravitational field of vacuum domain walls and strings}.
\bjtitle{Phys. Rev. D}
\bvolume{23},
\bfpage{852}--\blpage{857}
(\byear{1981})
\doiurl{10.1103/PhysRevD.23.852}
\end{barticle}
\endbibitem

%%% 8
\bibitem[\protect\citeauthoryear{Hong et~al.}{2023}]{hong2023spin}
\begin{barticle}
\bauthor{\bsnm{Hong}, \binits{D.}},
\bauthor{\bsnm{Lee}, \binits{J.}},
\bauthor{\bsnm{Kim}, \binits{J.}},
\bauthor{\bsnm{Jung}, \binits{J.H.}},
\bauthor{\bsnm{Lee}, \binits{K.}},
\bauthor{\bsnm{Kang}, \binits{S.}},
\bauthor{\bsnm{Shin}, \binits{Y.}}:
\batitle{Spin-driven stationary turbulence in spinor {B}ose-{E}instein
  condensates}.
\bjtitle{Phys. Rev. A}
\bvolume{108},
\bfpage{013318}
(\byear{2023})
\doiurl{10.1103/PhysRevA.108.013318}
\end{barticle}
\endbibitem

%%% 9
\bibitem[\protect\citeauthoryear{Ruben et~al.}{2010}]{ruben2010texture}
\begin{barticle}
\bauthor{\bsnm{Ruben}, \binits{G.}},
\bauthor{\bsnm{Morgan}, \binits{M.J.}},
\bauthor{\bsnm{Paganin}, \binits{D.M.}}:
\batitle{Texture control in a pseudospin {B}ose-{E}instein condensate}.
\bjtitle{Phys. Rev. Lett.}
\bvolume{105},
\bfpage{220402}
(\byear{2010})
\doiurl{10.1103/PhysRevLett.105.220402}
\end{barticle}
\endbibitem

%%% 10
\bibitem[\protect\citeauthoryear{Naraschewski and
  Glauber}{1999}]{naraschewski1999spatial}
\begin{barticle}
\bauthor{\bsnm{Naraschewski}, \binits{M.}},
\bauthor{\bsnm{Glauber}, \binits{R.J.}}:
\batitle{Spatial coherence and density correlations of trapped {B}ose gases}.
\bjtitle{Phys. Rev. A}
\bvolume{59},
\bfpage{4595}--\blpage{4607}
(\byear{1999})
\doiurl{10.1103/PhysRevA.59.4595}
\end{barticle}
\endbibitem

%%% 11
\bibitem[\protect\citeauthoryear{Henseler and
  Shapiro}{2008}]{henseler2008density}
\begin{barticle}
\bauthor{\bsnm{Henseler}, \binits{P.}},
\bauthor{\bsnm{Shapiro}, \binits{B.}}:
\batitle{Density correlations in cold atomic gases: Atomic speckles in the
  presence of disorder}.
\bjtitle{Phys. Rev. A}
\bvolume{77},
\bfpage{033624}
(\byear{2008})
\doiurl{10.1103/PhysRevA.77.033624}
\end{barticle}
\endbibitem

%%% 12
\bibitem[\protect\citeauthoryear{Kohnen and Nyman}{2015}]{kohnen2015temporal}
\begin{barticle}
\bauthor{\bsnm{Kohnen}, \binits{M.}},
\bauthor{\bsnm{Nyman}, \binits{R.A.}}:
\batitle{Temporal and spatiotemporal correlation functions for trapped {B}ose
  gases}.
\bjtitle{Phys. Rev. A}
\bvolume{91},
\bfpage{033612}
(\byear{2015})
\doiurl{10.1103/PhysRevA.91.033612}
\end{barticle}
\endbibitem

%%% 13
\bibitem[\protect\citeauthoryear{Jeltes et~al.}{2007}]{jeltes2007comparison}
\begin{barticle}
\bauthor{\bsnm{Jeltes}, \binits{T.}},
\bauthor{\bsnm{McNamara}, \binits{J.M.}},
\bauthor{\bsnm{Hogervorst}, \binits{W.}},
\bauthor{\bsnm{Vassen}, \binits{W.}},
\bauthor{\bsnm{Krachmalnicoff}, \binits{V.}},
\bauthor{\bsnm{Schellekens}, \binits{M.}},
\bauthor{\bsnm{Perrin}, \binits{A.}},
\bauthor{\bsnm{Chang}, \binits{H.}},
\bauthor{\bsnm{Boiron}, \binits{D.}},
\bauthor{\bsnm{Aspect}, \binits{A.}}, \betal:
\batitle{Comparison of the {H}anbury {B}rown--{T}wiss effect for bosons and
  fermions}.
\bjtitle{Nature}
\bvolume{445}(\bissue{7126}),
\bfpage{402}--\blpage{405}
(\byear{2007})
\doiurl{10.1038/nature05513}
\end{barticle}
\endbibitem

%%% 14
\bibitem[\protect\citeauthoryear{Perrin et~al.}{2012}]{perrin2012hanbury}
\begin{barticle}
\bauthor{\bsnm{Perrin}, \binits{A.}},
\bauthor{\bsnm{B{\"u}cker}, \binits{R.}},
\bauthor{\bsnm{Manz}, \binits{S.}},
\bauthor{\bsnm{Betz}, \binits{T.}},
\bauthor{\bsnm{Koller}, \binits{C.}},
\bauthor{\bsnm{Plisson}, \binits{T.}},
\bauthor{\bsnm{Schumm}, \binits{T.}},
\bauthor{\bsnm{Schmiedmayer}, \binits{J.}}:
\batitle{Hanbury {B}rown and {T}wiss correlations across the {B}ose-{E}instein
  condensation threshold}.
\bjtitle{Nature Physics}
\bvolume{8}(\bissue{3}),
\bfpage{195}--\blpage{198}
(\byear{2012})
\doiurl{10.1038/nphys2212}
\end{barticle}
\endbibitem

%%% 15
\bibitem[\protect\citeauthoryear{Hertkorn et~al.}{2021}]{hertkorn2021density}
\begin{barticle}
\bauthor{\bsnm{Hertkorn}, \binits{J.}},
\bauthor{\bsnm{Schmidt}, \binits{J.-N.}},
\bauthor{\bsnm{B\"ottcher}, \binits{F.}},
\bauthor{\bsnm{Guo}, \binits{M.}},
\bauthor{\bsnm{Schmidt}, \binits{M.}},
\bauthor{\bsnm{Ng}, \binits{K.S.H.}},
\bauthor{\bsnm{Graham}, \binits{S.D.}},
\bauthor{\bsnm{B\"uchler}, \binits{H.P.}},
\bauthor{\bsnm{Langen}, \binits{T.}},
\bauthor{\bsnm{Zwierlein}, \binits{M.}},
\bauthor{\bsnm{Pfau}, \binits{T.}}:
\batitle{Density fluctuations across the superfluid-supersolid phase transition
  in a dipolar quantum gas}.
\bjtitle{Phys. Rev. X}
\bvolume{11},
\bfpage{011037}
(\byear{2021})
\doiurl{10.1103/PhysRevX.11.011037}
\end{barticle}
\endbibitem

%%% 16
\bibitem[\protect\citeauthoryear{F{\"o}lling}{}]{folling2014quantum}
\begin{botherref}
\oauthor{\bsnm{F{\"o}lling}, \binits{S.}}:
Chapter 8.
Quantum Noise Correlation Experiments with Ultracold Atoms,
pp. 145--177.
\doiurl{10.1142/9781783264766_0008}
\end{botherref}
\endbibitem

%%% 17
\bibitem[\protect\citeauthoryear{Nagler et~al.}{2022}]{nagler2022observing}
\begin{barticle}
\bauthor{\bsnm{Nagler}, \binits{B.}},
\bauthor{\bsnm{Barbosa}, \binits{S.}},
\bauthor{\bsnm{Koch}, \binits{J.}},
\bauthor{\bsnm{Orso}, \binits{G.}},
\bauthor{\bsnm{Widera}, \binits{A.}}:
\batitle{Observing the loss and revival of long-range phase coherence through
  disorder quenches}.
\bjtitle{Proceedings of the National Academy of Sciences}
\bvolume{119}(\bissue{1}),
\bfpage{2111078118}
(\byear{2022})
\doiurl{10.1073/pnas.2111078118}
\end{barticle}
\endbibitem

%%% 18
\bibitem[\protect\citeauthoryear{Middleton-Spencer
  et~al.}{2023}]{middleton2023strong}
\begin{barticle}
\bauthor{\bsnm{Middleton-Spencer}, \binits{H.A.J.}},
\bauthor{\bsnm{Orozco}, \binits{A.D.G.}},
\bauthor{\bsnm{Galantucci}, \binits{L.}},
\bauthor{\bsnm{Moreno}, \binits{M.}},
\bauthor{\bsnm{Parker}, \binits{N.G.}},
\bauthor{\bsnm{Machado}, \binits{L.A.}},
\bauthor{\bsnm{Bagnato}, \binits{V.S.}},
\bauthor{\bsnm{Barenghi}, \binits{C.F.}}:
\batitle{Strong quantum turbulence in {B}ose-{E}instein condensates}.
\bjtitle{Phys. Rev. Res.}
\bvolume{5},
\bfpage{043081}
(\byear{2023})
\doiurl{10.1103/PhysRevResearch.5.043081}
\end{barticle}
\endbibitem

%%% 19
\bibitem[\protect\citeauthoryear{Barenghi et~al.}{2023}]{barenghi2023types}
\begin{barticle}
\bauthor{\bsnm{Barenghi}, \binits{C.F.}},
\bauthor{\bsnm{Middleton-Spencer}, \binits{H.A.J.}},
\bauthor{\bsnm{Galantucci}, \binits{L.}},
\bauthor{\bsnm{Parker}, \binits{N.G.}}:
\batitle{Types of quantum turbulence}.
\bjtitle{AVS Quantum Science}
\bvolume{5}(\bissue{2}),
\bfpage{025601}
(\byear{2023})
\doiurl{10.1116/5.0146107}
\end{barticle}
\endbibitem

%%% 20
\bibitem[\protect\citeauthoryear{Caracanhas et~al.}{2012}]{caracanhas2012self}
\begin{barticle}
\bauthor{\bsnm{Caracanhas}, \binits{M.}},
\bauthor{\bsnm{Fetter}, \binits{A.L.}},
\bauthor{\bsnm{Muniz}, \binits{S.R.}},
\bauthor{\bsnm{Magalhães}, \binits{K.M.F.}},
\bauthor{\bsnm{Roati}, \binits{G.}},
\bauthor{\bsnm{Bagnato}, \binits{G.}},
\bauthor{\bsnm{Bagnato}, \binits{V.S.}}:
\batitle{Self-similar expansion of the density profile in a turbulent
  {B}ose-{E}instein condensate}.
\bjtitle{Journal of Low Temperature Physics}
\bvolume{166},
\bfpage{49}--\blpage{58}
(\byear{2012})
\doiurl{10.1007/s10909-011-0409-2}
\end{barticle}
\endbibitem

%%% 21
\bibitem[\protect\citeauthoryear{Caracanhas et~al.}{2013}]{caracanhas2013self}
\begin{barticle}
\bauthor{\bsnm{Caracanhas}, \binits{M.}},
\bauthor{\bsnm{Fetter}, \binits{A.}},
\bauthor{\bsnm{Baym}, \binits{G.}},
\bauthor{\bsnm{Muniz}, \binits{S.R.}},
\bauthor{\bsnm{Bagnato}, \binits{V.S.}}:
\batitle{Self-similar expansion of a turbulent {B}ose-{E}instein condensate: A
  generalized hydrodynamic model}.
\bjtitle{Journal of Low Temperature Physics}
\bvolume{170},
\bfpage{133}--\blpage{142}
(\byear{2013})
\doiurl{10.1007/s10909-012-0776-3}
\end{barticle}
\endbibitem

%%% 22
\bibitem[\protect\citeauthoryear{Thompson et~al.}{2013}]{thompson2013evidence}
\begin{barticle}
\bauthor{\bsnm{Thompson}, \binits{K.J.}},
\bauthor{\bsnm{Bagnato}, \binits{G.G.}},
\bauthor{\bsnm{Telles}, \binits{G.D.}},
\bauthor{\bsnm{Caracanhas}, \binits{M.A.}},
\bauthor{\bsnm{Santos}, \binits{F.E.A.}},
\bauthor{\bsnm{Bagnato}, \binits{V.S.}}:
\batitle{Evidence of power law behavior in the momentum distribution of a
  turbulent trapped {B}ose–{E}instein condensate}.
\bjtitle{Laser Physics Letters}
\bvolume{11}(\bissue{1}),
\bfpage{015501}
(\byear{2013})
\doiurl{10.1088/1612-2011/11/1/015501}
\end{barticle}
\endbibitem

%%% 23
\bibitem[\protect\citeauthoryear{Daniel García-Orozco
  et~al.}{2020}]{garcia2020intrascales}
\begin{barticle}
\bauthor{\bsnm{Daniel~García-Orozco}, \binits{A.}},
\bauthor{\bsnm{Madeira}, \binits{L.}},
\bauthor{\bsnm{Galantucci}, \binits{L.}},
\bauthor{\bsnm{Barenghi}, \binits{C.F.}},
\bauthor{\bsnm{Bagnato}, \binits{V.S.}}:
\batitle{Intra-scales energy transfer during the evolution of turbulence in a
  trapped {B}ose-{E}instein condensate(a)}.
\bjtitle{Europhysics Letters}
\bvolume{130}(\bissue{4}),
\bfpage{46001}
(\byear{2020})
\doiurl{10.1209/0295-5075/130/46001}
\end{barticle}
\endbibitem

%%% 24
\bibitem[\protect\citeauthoryear{Garc\'{\i}a-Orozco
  et~al.}{2022}]{garcia2022universal}
\begin{barticle}
\bauthor{\bsnm{Garc\'{\i}a-Orozco}, \binits{A.D.}},
\bauthor{\bsnm{Madeira}, \binits{L.}},
\bauthor{\bsnm{Moreno-Armijos}, \binits{M.A.}},
\bauthor{\bsnm{Fritsch}, \binits{A.R.}},
\bauthor{\bsnm{Tavares}, \binits{P.E.S.}},
\bauthor{\bsnm{Castilho}, \binits{P.C.M.}},
\bauthor{\bsnm{Cidrim}, \binits{A.}},
\bauthor{\bsnm{Roati}, \binits{G.}},
\bauthor{\bsnm{Bagnato}, \binits{V.S.}}:
\batitle{Universal dynamics of a turbulent superfluid {B}ose gas}.
\bjtitle{Physical Review A}
\bvolume{106},
\bfpage{023314}
(\byear{2022})
\doiurl{10.1103/PhysRevA.106.023314}
\end{barticle}
\endbibitem

%%% 25
\bibitem[\protect\citeauthoryear{Moreno-Armijos
  et~al.}{2025}]{moreno2025observation}
\begin{barticle}
\bauthor{\bsnm{Moreno-Armijos}, \binits{M.A.}},
\bauthor{\bsnm{Fritsch}, \binits{A.R.}},
\bauthor{\bsnm{Garc\'{\i}a-Orozco}, \binits{A.D.}},
\bauthor{\bsnm{Sab}, \binits{S.}},
\bauthor{\bsnm{Telles}, \binits{G.}},
\bauthor{\bsnm{Zhu}, \binits{Y.}},
\bauthor{\bsnm{Madeira}, \binits{L.}},
\bauthor{\bsnm{Nazarenko}, \binits{S.}},
\bauthor{\bsnm{Yukalov}, \binits{V.I.}},
\bauthor{\bsnm{Bagnato}, \binits{V.S.}}:
\batitle{Observation of relaxation stages in a nonequilibrium closed quantum
  system: Decaying turbulence in a trapped superfluid}.
\bjtitle{Phys. Rev. Lett.}
\bvolume{134},
\bfpage{023401}
(\byear{2025})
\doiurl{10.1103/PhysRevLett.134.023401}
\end{barticle}
\endbibitem

%%% 26
\bibitem[\protect\citeauthoryear{Navon et~al.}{2016}]{navon2016emergence}
\begin{barticle}
\bauthor{\bsnm{Navon}, \binits{N.}},
\bauthor{\bsnm{Gaunt}, \binits{A.L.}},
\bauthor{\bsnm{Smith}, \binits{R.P.}},
\bauthor{\bsnm{Hadzibabic}, \binits{Z.}}:
\batitle{Emergence of a turbulent cascade in a quantum gas}.
\bjtitle{Nature}
\bvolume{539},
\bfpage{72}--\blpage{75}
(\byear{2016})
\doiurl{10.1038/nature20114}
\end{barticle}
\endbibitem

%%% 27
\bibitem[\protect\citeauthoryear{Fabbri et~al.}{2011}]{fabbri2011momentum}
\begin{barticle}
\bauthor{\bsnm{Fabbri}, \binits{N.}},
\bauthor{\bsnm{Cl\'ement}, \binits{D.}},
\bauthor{\bsnm{Fallani}, \binits{L.}},
\bauthor{\bsnm{Fort}, \binits{C.}},
\bauthor{\bsnm{Inguscio}, \binits{M.}}:
\batitle{Momentum-resolved study of an array of one-dimensional strongly
  phase-fluctuating {B}ose gases}.
\bjtitle{Phys. Rev. A}
\bvolume{83},
\bfpage{031604}
(\byear{2011})
\doiurl{10.1103/PhysRevA.83.031604}
\end{barticle}
\endbibitem

%%% 28
\bibitem[\protect\citeauthoryear{Pethick and Smith}{2008}]{pethick2008bose}
\begin{bbook}
\bauthor{\bsnm{Pethick}, \binits{C.J.}},
\bauthor{\bsnm{Smith}, \binits{H.}}:
\bbtitle{Bose–{E}instein Condensation in Dilute Gases},
\bedition{2}nd edn.
\bpublisher{Cambridge University Press},
\blocation{Cambridge}
(\byear{2008})
\end{bbook}
\endbibitem

%%% 29
\bibitem[\protect\citeauthoryear{Ketterle et~al.}{1999}]{ketterle1999making}
\begin{bchapter}
\bauthor{\bsnm{Ketterle}, \binits{W.}},
\bauthor{\bsnm{Durfee}, \binits{D.S.}},
\bauthor{\bsnm{Stamper-Kurn}, \binits{D.}}:
\bctitle{Making, probing and understanding {B}ose-{E}instein condensates}.
In: \bbtitle{Bose-{E}instein Condensation in Atomic Gases},
pp. \bfpage{67}--\blpage{176}.
\bpublisher{IOS Press},
\blocation{Varenna}
(\byear{1999})
\end{bchapter}
\endbibitem

%%% 30
\bibitem[\protect\citeauthoryear{Chiofalo et~al.}{2000}]{itime}
\begin{barticle}
\bauthor{\bsnm{Chiofalo}, \binits{M.L.}},
\bauthor{\bsnm{Succi}, \binits{S.}},
\bauthor{\bsnm{Tosi}, \binits{M.P.}}:
\batitle{Ground state of trapped interacting {B}ose-{E}instein condensates by
  an explicit imaginary-time algorithm}.
\bjtitle{Phys. Rev. E}
\bvolume{62},
\bfpage{7438}--\blpage{7444}
(\byear{2000})
\doiurl{10.1103/PhysRevE.62.7438}
\end{barticle}
\endbibitem

\end{thebibliography}

\end{document}